# Landau Level Physics in a Quantum Well: new singular features in Magnetization and violations of de Haas – van Alphen periodicities

Georgios Konstantinou and Konstantinos Moulopoulos

*University of Cyprus, Department of Physics, 1678 Nicosia, Cyprus*

**Abstract.** Analytical calculations based on a Landau Level (LL) picture are reported for an interface (with a finite-width Quantum Well (QW)) and for a fully three-dimensional charged quantum electronic system in an external magnetic field. They lead to a sequence of previously unnoticed singular features in global magnetization and magnetic susceptibility that lead to nontrivial corrections to the standard de Haas - van Alphen periods. Additional features due to Zeeman splitting are also reported (such as new energy minima that originate from the interplay of QW, Zeeman and LL Physics) that are possibly useful for the design of quantum devices. A corresponding calculation in a Composite Fermion picture leads to new predictions on magnetic response properties of a fully-interacting electron liquid in a finite-width interface.

## I. INTRODUCTION - MOTION IN TWO DIMENSIONS

Let us start with a 2-dimensional (2D) system of noninteracting electrons (each with charge -e, effective mass m and spin s=1/2) moving in an external magnetic field **B** perpendicular to the system, and with no Zeeman splitting (i.e. $g^* = 0$). The single-electron energy is[1]:

$$\varepsilon_n = \left(n + \tfrac{1}{2}\right)\hbar\omega_c, \ n = 0,1,2... \text{ with } \omega_c = \frac{|eB|}{mc}$$

The total Energy per electron (in units of 2D Fermi energy) and Magnetization can be determined in closed form and they are shown in **Fig. 1** ($\Phi_o = hc/e$ is the flux quantum):

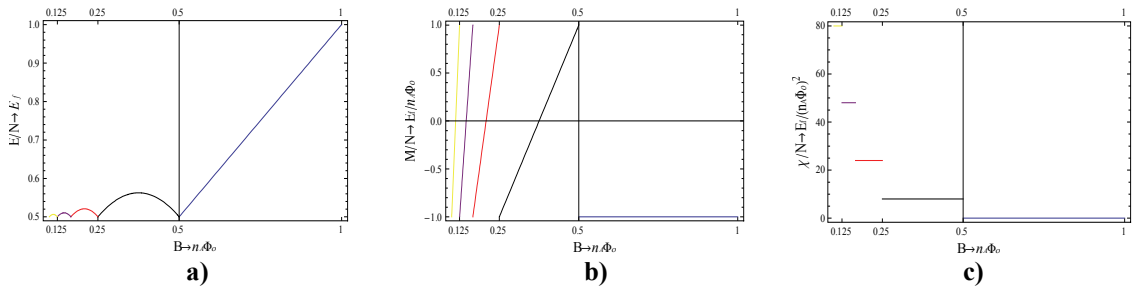

**FIGURE 1:** **a)** Energy, **b)** Magnetization and **c)** Susceptibility per electron as functions of **B**.

# II. MOTION IN QUASI –2D (INTERFACE)

Let us now consider an interface, and carefully take into account the nonzero width **d** as an essential variable in the problem: Here, it is advantageous to keep **B** fixed (but within predescribed ranges of values, so that the filling factor is constant) and vary **d**.

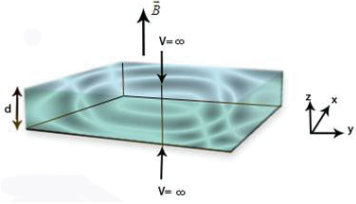

**FIGURE 2:** Example of the space in which the particles (electrons) live. Let's suppose that we turn on an infinite scalar potential at the two macroscopic surfaces (north and south).

The single electron energy (for $g^* = 0$) is now:

$$\varepsilon_n = \left(n + \tfrac{1}{2}\right)\hbar\omega_c + \frac{\hbar^2 \pi^2 n_z^2}{2md^2}, \quad n_z = 1, 2..$$

From the point of view of a single electron (namely, what state it should occupy so that the total Energy is the lowest possible) we notice that there are *points of decision* that are described by equalities of single-particle energies[2]. So, i.e. in **Fig. 3** when **B** varies between the values: $\tfrac{1}{6} n_A \Phi_o \leq B \leq \tfrac{1}{4} n_A \Phi_o$, where $n_A$ is the areal density, $n_A = N/S$, we have (by increasing **d** continuously):

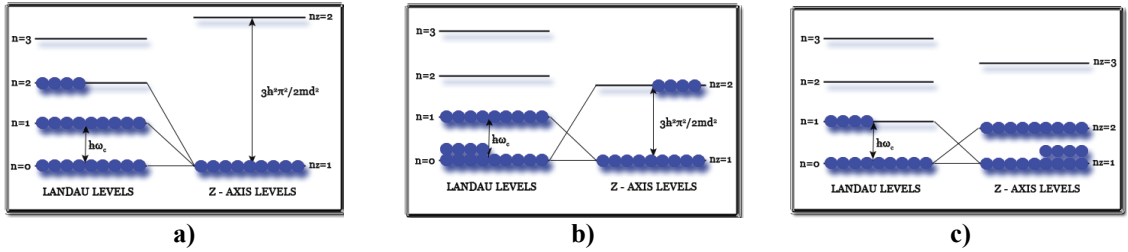

a)                                       b)                                       c)

**FIGURE 3:** In **a)**, when **d** is very small, the system occupies only LLs (because the energy gap between the QW levels is very large). In **b)**, **d** is sufficiently large so that the system "prefers" the occupation of the second QW level. In **c)**, the second QW level is completely filled with electrons, while the LL n=1 is partially filled due to the very large **d**.

When the occupied levels are found, we can write down the total Energy for each range of values of **d** and of **B** and the results for $n_A = 10^{16} \ m^{-2}$ are shown in **Fig. 4, 5, 6** below:

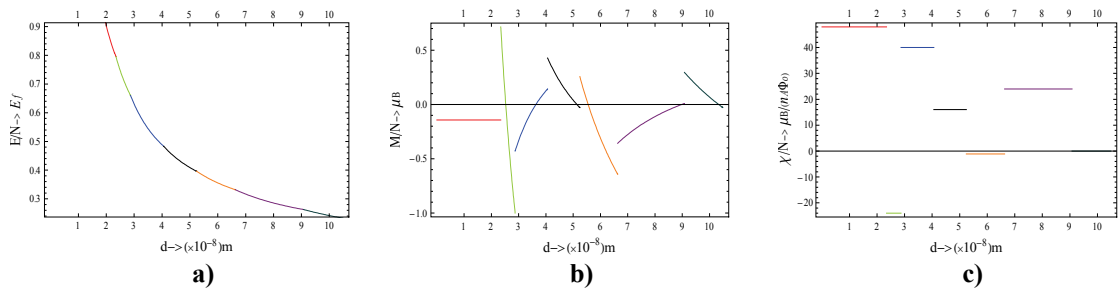

a)                                       b)                                       c)

**FIGURE 4: a)** Energy, **b)** Magnetization and **c)** Susceptibility as functions of **d**, when **B** is: $\tfrac{1}{7} n_A \Phi_o$.

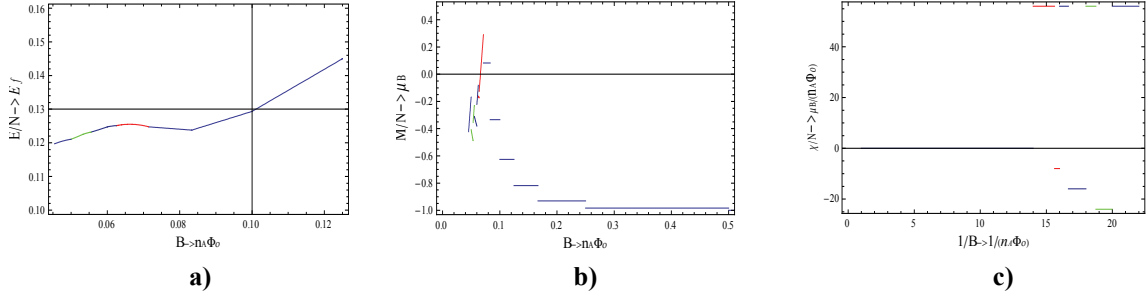

**FIGURE 5: a)** Energy, **b)** Magnetization and **c)** Susceptibility as functions of **B** (or **1/B**), when **d** is 242 nm. (These discontinuities in magnetization are violations of the standard de Haas - van Alphen periodicities for a 2D system).

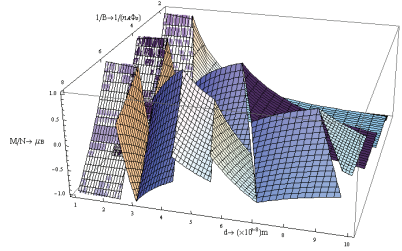

**FIGURE 6:** Magnetization as a function of both **B** and **d.**

## III. MOTION IN FULL 3D SPACE (ZEEMAN SPLIT OFF)

Let us now allow the system to move in the full 3D space[3]. In this case, the single fermion energy reads:

$$\varepsilon_n = \left(n + \tfrac{1}{2}\right)\hbar\omega_c + \frac{\hbar^2 k_z^2}{2m}, \quad k_z: \text{continuous, or } k_z = \frac{2\pi l}{L}, \quad l = 0, \pm 1, \pm 2.., \quad L \to \infty$$

It turns out that we now have an 1D Fermi segment along the z-direction (for each LL): $k_f = \pi^2 l_B^2 n$, with $l_B$ the magnetic length ($=\sqrt{\hbar c / eB}$) and $n$ the 3-D density. The number of $k_f$'s (hence of LLs) defines the ranges of **B** values. The points that these ranges intersect give the exact quantal corrections of the standard semiclassical de Haas - van Alphen periodicities in 3D. As we move to the left we asymptotically recover the de Haas - van Alphen behaviour[4].

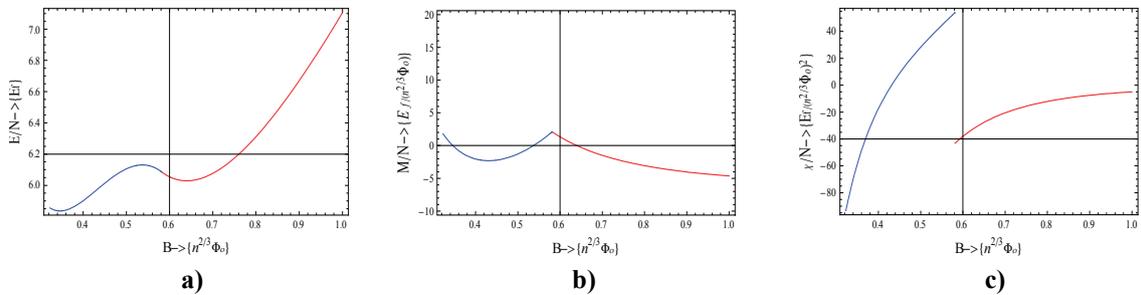

**FIGURE 7: a)** Energy, **b)** Magnetization and **c)** Susceptibilty as functions of **B**.

## IV. INCLUDING ZEEMAN SPLITTING IN ALL CALCULATIONS

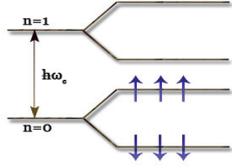

We can show[2] that the thermodynamic properties as functions of the gyromagnetic ratio $g^*$ can change dramatically (i.e see **Fig. 8**). The single fermion energy, which takes into account the spin interaction with the magnetic field, is as follows:

$$\varepsilon_n = \left(n + \frac{1}{2} \pm \frac{g^*}{4}\frac{m^*}{m}\right)\hbar\omega_c + \frac{\hbar^2 k_z^2}{2m},$$ where $m^*$ is the electron's effective mass.

### Motion in full 3D space

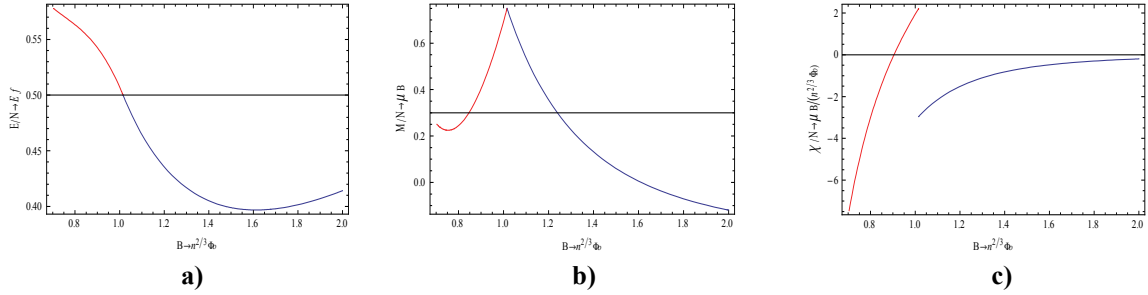

**FIGURE 8: a)** Energy, **b)** Magnetization, and **c)** Susceptibility as functions of the magnetic field when $g^* = 1.5$. (*The minimum in Energy* should be noted, when $g^*$ is sufficiently large).

## V. COMPOSITE FERMIONS (FOR $g^* = 0$)

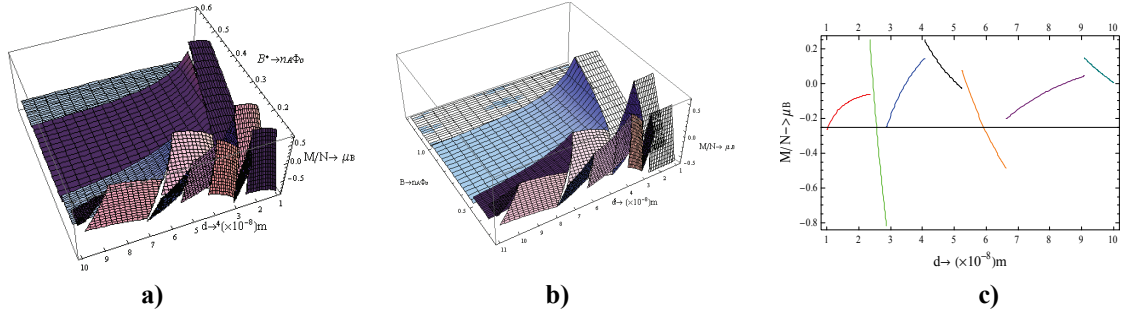

**FIGURE 9: a)** Magnetization (of the fully-interacting system) as a function of the *effective* magnetic field[5] and width. **b):** Magnetization (of the noninteracting system) as a function of the magnetic field and width. **c):** Magnetization per Composite Fermion as a function of the width, when the *effective* magnetic field ($B^* = B - 2pn_A\Phi_o$) is: $\frac{1}{7}n_A\Phi_o$ (and $p = 1$).

## VI. CONCLUSIONS

In all the above, for interfaces and for full 3D space, we have obtained *exact corrections* to the standard de Haas – van Alphen periodicities and *new signatures* (in magnetic properties) of electron-electron interactions (compare **Fig. 9c** with **Fig. 4b**).